# The Necessity for Hardware QoS Support for Server Consolidation and Cloud Computing

Javier Merino, Valentin Puente, José Ángel Gregorio

University of Cantabria, Spain

**Abstract.** *Chip multiprocessors (CMPs) are ubiquitous in most of today's computing fields. Although they provide noticeable benefits in terms of performance, cost and power efficiency, they also introduce some new issues. In this paper we analyze how the interference from Virtual Private Servers running in other cores is a significant component of performance unpredictability and can threaten the attainment of cloud computing. Even if virtualization is used, the sharing of the on-chip section of the memory hierarchy by different cores makes performance isolation strongly dependent on what is running elsewhere in the system. We will show in three actual computing systems, based on Sun UltraSparc T1, Sun UltraSparc T2 and Intel Xeon processors, how state-of-the-art virtualization techniques are unable to guarantee performance isolation in a representative workload such as SPECweb2005. In an especially conceived near worst-case scenario, it is possible to reduce the performance achieved by a Solaris Zones consolidated server for this suite of benchmarks in a Sun Fire T1000 and a Sun Enterprise T5120 by up to 80%. The performance drop observed by a Xen consolidated server running in a HP Proliant DL160 G5 is almost 45%. For all systems under study, off-chip bandwidth is shown to be the most critical resource.*

**Keywords:** Server Consolidation, CQoS, performance isolation, CMP, off-chip bandwidth

## 1 Introduction

Server consolidation is a first order issue in today's computing infrastructures and services. The combination of a rising number of cores per chip and virtualization techniques makes it feasible to condense a higher number of services into a smaller set of physical computing resources, thus promoting the spread of cloud computing. Chip multi-processors (CMPs) facilitate the consolidation process, significantly reducing the total cost of ownership of computing infrastructure. Next generation multi-core CMPs will increase the number of services running per chip. Currently in multi-computing, such as blade enclosures or multi-chip multiprocessors, virtualization is the most important tool to help in the deployment of such infrastructures. When operating such services, for security and performance reasons it is advisable to isolate and regulate services and virtual machines, through the definition of Virtual Private Servers (VPSs). VPSs are the best candidate to provide complete isolation and fit computing requirements to the available hardware resources. A cloud computing provider, with a software framework such as Eucalyptus [16], owns clusters of multi-core CPUs and offers VPSs to their customers. This concept benefits both parties: the provider can fully exploit its servers obtaining income from them, and the customers have servers at their disposal without worrying about hardware failures, electricity, air-conditioning, etc. The VPSs are usually implemented using some form of virtualization, such as Xen [3], OpenVZ [21] or Solaris Zones [22]. These virtualization techniques attempt to provide performance isolation to the underlying virtual machines.

In most cases, VPS performance is bounded by assigning a limited number of cores to each of them, with a specific amount of physical memory, hard disk space, restricting the network bandwidth availability, etc. Using such features, it is possible to regulate access to the hardware for all the services, guaranteeing that the total running VPS will not overrun the available resources

Today's data centers use CMPs extensively in their computing infrastructure. Those CMPs share some resources of the system where they are physically allocated. On these data centers there may be thousands of VPSs, each of them expecting performance isolation. For example, a cloud computing provider may own several servers with a Niagara chip (8 cores) [20] and customers can rent $n$-core VPS. Unfortunately, when CMPs are used as the hardware platform for VPS deployment, software techniques have limited on-chip resource management capabilities. The OS or hypervisor knows the performance requirements but is unable to properly regulate the access to resources, such as part of the CMP memory hierarchy. Those resources are too fine-grained to be controlled through a software approach like virtualization. This could potentially imply that a badly behaving VPS can harm the performance of other VPSs running in the same CMP. Therefore, the service provider cannot realistically sign a Service Level Agreement (SLA) which guarantees a certain Quality of Service. The current solution, separating VPSs in resources that do not share any section of the memory hierarchy (separate chips) does not scale and is not cost effective: many-core CMPs will condense tens of processing cores, and the only reasonable approach is to share part of the CMP among different VPSs.

In this work we will show how it is possible, in a state-of-the-art environment, to observe destructive interference between VPSs. In particular, uncontrolled access to shared resources in the memory hierarchy by one VPS could significantly degrade the performance of other VPSs running in the same chip. In order to know the extent of this degradation we define a worst-case scenario where the offending VPS is running applications that try to monopolize all on-chip resources and exhaust off-chip bandwidth. Combining this situation with a VPS running a workload suitable for consolidation such as SpecWEB2005, a massive amount of performance loss is observed. Although this scenario is not necessarily common, it could be possible due to malicious intentions of some VPS users. We ran these experiments using Sun UltraSparc T1- and UltraSparc T2-based Servers and an Intel Xeon Harpertown-based 8-core server running Solaris and GNU/Linux Operating Systems, respectively. The virtualization techniques used are Solaris Zones[22] and Xen[3], respectively.

As a result of these experiments this work makes the following main contributions:

- Current solutions for Virtual Private Servers are not enough to isolate performance in consolidated servers. Virtualization software does not have the necessary mechanisms available to guarantee performance stability. In all systems, the CMP resource with biggest impact on performance interference is shown to be the off-chip bandwidth.

- With currently available CMP commercial hardware, it is not possible to guarantee a given Quality of Service, i.e. commercial CMPs do not provide the hardware capabilities to distinguish the utilization of shared resources by different VPS and therefore they are unable to prioritize utilization or to enable the virtualization layer to do so.

- A VPS should indicate to the underlying hardware the priorities of the workloads running. In order to guarantee fairness, hardware should supply the necessary mechanisms to balance the utilization of shared resources according to VPS requirements.

Although, as is indicated in Section 5, there is a significant amount of previous work analyzing these problems, even proposing effective solutions to alleviate them, to our knowledge this is the first work that shows the extent of the problem over real platforms running a state-of-the-art software stack.

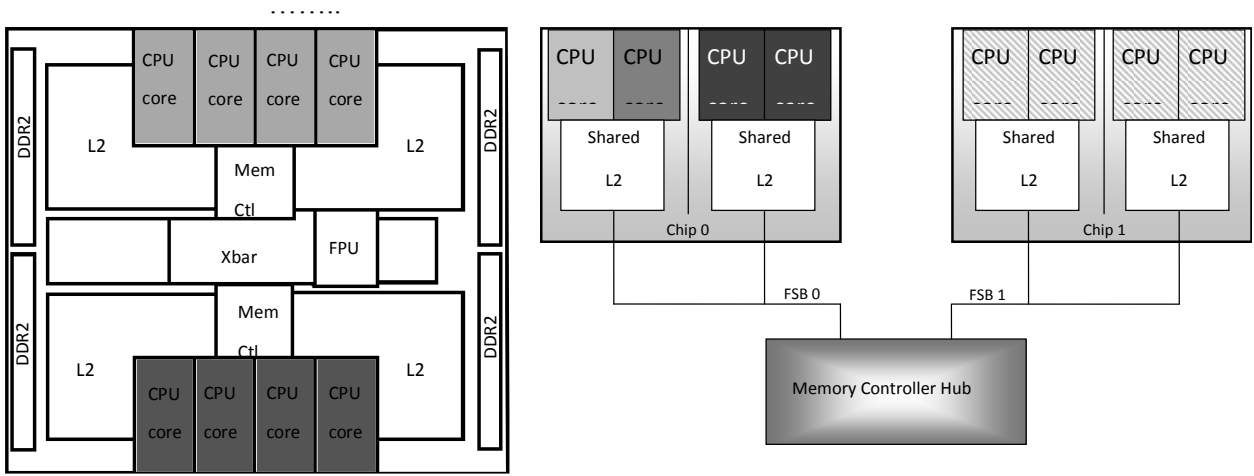

***Fig. 1*** *(left) Sun Fire T1000 processor and memory controller floor plan, (right) HP DL160 G5 processor and memory controller floor plan*

The rest of the paper is organized as follows: Section 2 describes the experimental set-up. In section 3 a harmful workload, capable of generating inter VPS interference is defined and analyzed with a well-known paper-and-pencil multithread application. Section 4 quantifies the effect of the previously defined application in a consolidated server running a realistic workload and discusses the results obtained. Finally, Section 5 reviews related work and Section 6 states the main conclusions of the paper.

## 2 Experimental Set-up

### 2.1 Basic System Configuration

We selected two possible hardware configurations to evaluate the capabilities of server consolidation in state-of-the-art system setups. The main hardware and basic software characteristics of both platforms are summarized in **Table 1**. The Sun Fire T1000 is a highly efficient system, especially suitable for throughput computing. The HP Proliant DL160 utilization is more flexible due to the higher capabilities of the processor. In the case of the Sun Fire T1000, we divide this 8-core single-chip server (Fig1.left) into two 4-core VPS using Solaris Zones [22]. We distinguish these zones by calling them *primary* and *secondary*. Using the tools available for resource management in Solaris Zones, we assign 16 logical CPUs (equivalent to 4 physical ones) to the primary zone and 15 logical CPUs to the secondary zone. The last thread of the last core is reserved for the global zone, which runs the non-virtualized OS and remains idle during the measurements. The Sun Enterprise T5120 uses an updated version of this processor which is structurally similar to the UltraSparc T1. Additionally, the storage subsystem and networking is significantly better than in Sun Fire T1000.

The HP Proliant DL160, as can be appreciated in Fig 1(right), is a multi-chip system. In this case it is useful to widen the set of configurations chosen for each VPS. We used Xen [3] as the virtualization platform. Using different configurations and locations for the primary and secondary VPS we can analyze the differences between inter-chip or intra-chip sharing effects. We will mix VPSs running in the same die and same chip.

|  | **Sun Fire T1000** | **Sun Enterprise T5120** | **HP Proliant DL160 G5** |
|---|---|---|---|
| **Processor Model** | Sun UltraSparc T1 | Sun UltraSparc T2 | Intel Xeon X5472 (Harpertown) with Intel 5400 Chipset |
| **Threads, Cores, dies, chips** | 32, 8, 1, 1 | 32, 4,1,1 | 8, 8, 4, 2 |
| **L1I** | Private, 16KB 4-way set associative, 32-byte lines, 1-per core | Private, 16KB 4-way set associative, 32-byte lines, 1-per core | Private, 32KB, 8-way set associative, 64-byte lines, 1-per core |
| **L1D** | Private, 8KB 4-way set associative, 16-byte lines, write-through, 1 per core | Private, 8KB 4-way set associative, 16-byte lines, write-through, 1 per core | Private, 32KB, 8-way set associative, 64-byte lines, 1-per core |
| **L2** | 3MB Shared per chip, 12-way set associative, 64-byte lines, 4 banks | 4MB Shared per chip, 16-way set associative, 64-byte lines, 8 banks | 6MB Shared per die, 24-way set associative, 64 byte lines |
| **Memory** | 8GB, 533Mhz, DDR2 | 8GB, 667Mhz, FB-DIMM | 16GB, 667Mhz FB-DIMM |
| **Operating System** | Solaris 10 | Solaris 10 | GNU/Linux Debian 5.0 |
| **Hard disk** | 1 SATA | 2 SAS RAID 0 | 2 SATA RAID 0 |
| **NIC** | Dual Gigabit Ethernet (bounded) | Quad Gigabit Ethernet (bounded) | Dual Gigabit Ethernet (bounded) |
| **Virtualization Software** | Solaris Zones | Solaris Zones | Xen v 3.2.1 |
| **Apache version** | Apache Tomcat 4.1 | Apache Tomcat 4.1 | Apache Tomcat 5.5 |

*Table 1. Systems configuration*

## 2.2 Workload Set-up

As the service to be consolidated, we chose a commercial workload that reflects a very common scenario where the hardware considered is used. We install Apache Tomcat [1] in the primary zone and evaluate its performance using the SPECweb2005 suite [23], with different states for the secondary VPS:

- **Idle**: The secondary VPS is executing nothing.

- **Running a Harmful Synthetic Application**: A program designed specifically to stress the memory hierarchy is running in the secondary VPS. Section 3 analyzes how it works.

SPECweb2005 clients and its backend simulator are run on another system, connected to the server through bounded Gigabit Ethernet adaptors. The web server dynamically generates a response using Java Server Pages (JSP), since we are using Apache Tomcat.

Our results show the number of simultaneous clients that passed the quality tests of the SPECweb2005 suite. That is, more than 95% of their requests finish in less than `TIME_GOOD` seconds and more than 99% finish in less than `TIME_TOLERABLE` seconds (both of them being parameters of the suite). The SPECweb2005 suite consists of three benchmarks:

- **Bank**: The bank benchmark simulates users checking their bank accounts and doing financial operations. As bank data is very sensitive, all requests are served using SSL. That means that each request needs more CPU than with other benchmarks.

- **E-Commerce:** This benchmark simulates the kind of web traffic received in an on-line store. It has two different phases: selecting what to buy and paying for that product. On the one hand, when browsing products, requests imply large unencrypted data transfers (there are a substantial number of images per page). On the other hand, when paying, requests are much smaller, but the traffic uses SSL in order to protect sensitive customer information.

- **Support:** In this benchmark, the simulated users browse a web site and end up downloading some software. Therefore, all traffic goes unencrypted and most of the requests are rather small, except for the download requests, which consist of various Mbytes. Even though the download requests are big, they do not impose a large burden on the web server since it is the only type of request that serves static content.

## 3    Harmful Synthetic Applications

The synthetic workloads are specially designed to stress the memory hierarchy of the systems under study. This load will allow us to establish a near to worst-case interference among different VPSs. This benchmark spawns as many threads as hardware contexts are available. Each thread creates an array of *s* bytes and reads or writes it using a stride of *j* bytes. For each hardware configuration and desired interference we tune parameters *j* and *s*. We will adjust those parameters to stress different shared stretches of the CMP memory hierarchy.

## 3.1 Sun UltraSparc T1 and T2

In these systems, the resources shared between cores in the memory hierarchy are the memory controllers, the L2 cache and its access through the internal crossbar. Therefore, we focused on adjusting two versions of the harmful application trying to find a near to worst-case scenario. We analyzed the systems under study running simple applications on the primary VPS and different variations of the harmful workload. The variations considered attempted to find a worst-case interference in on-chip shared resources and off-chip bandwidth separately. To maximize off-chip interference we developed a micro-kernel to obtain a high number of on-chip misses on all memory controllers but with a minimal number of the L2 sets monopolized. We will identify this kernel as *harm.off-chip*. It is much harder to develop a micro-kernel with only on-chip interference because if we want to monopolize on-chip resources, some off-chip bandwidth has to be used. In any case, we defined a workload to hog L2 capacity and on-chip bandwidth (at crossbar or bank level) reducing interference in memory controllers as much as possible. We denominate this micro-kernel as *harm.on-chip*.

The *harm.off-chip* micro-kernel was configured with 15[1] threads (1 per available hardware context in the secondary VPS), with a private array of 2MBytes per thread and stride of 256Kbytes in both systems[2]. The *harm.on-chip* micro-kernel was configured with 15 threads walking through a shared array of 3MB in UltraSparc T1 and 4MB in UltraSparc T2, with a stride of 64Bytes and a thread-dependent starting point. Given the fact that the cache is 12-way set associative in UltraSparc T1 and 16-way set associative in UltraSparc T2, each thread accesses every block in each set evicting every VPS block previously stored.

---

[1] The spare hardware context is devoted to the native OS and the virtualization layer.

[2] The L2 of the UltraSparc T1 processor has four banks, with the bank selection based on the physical address bits 7:6[20]. Thus, each bank has 1024 sets and the set is chosen from bits 17:8 of the physical address. Using a stride of 256Kbytes ($2^{6+2+10}$) performs all the accesses in the same set miss. The same reasoning is followed for the UltraSparc T2 system.

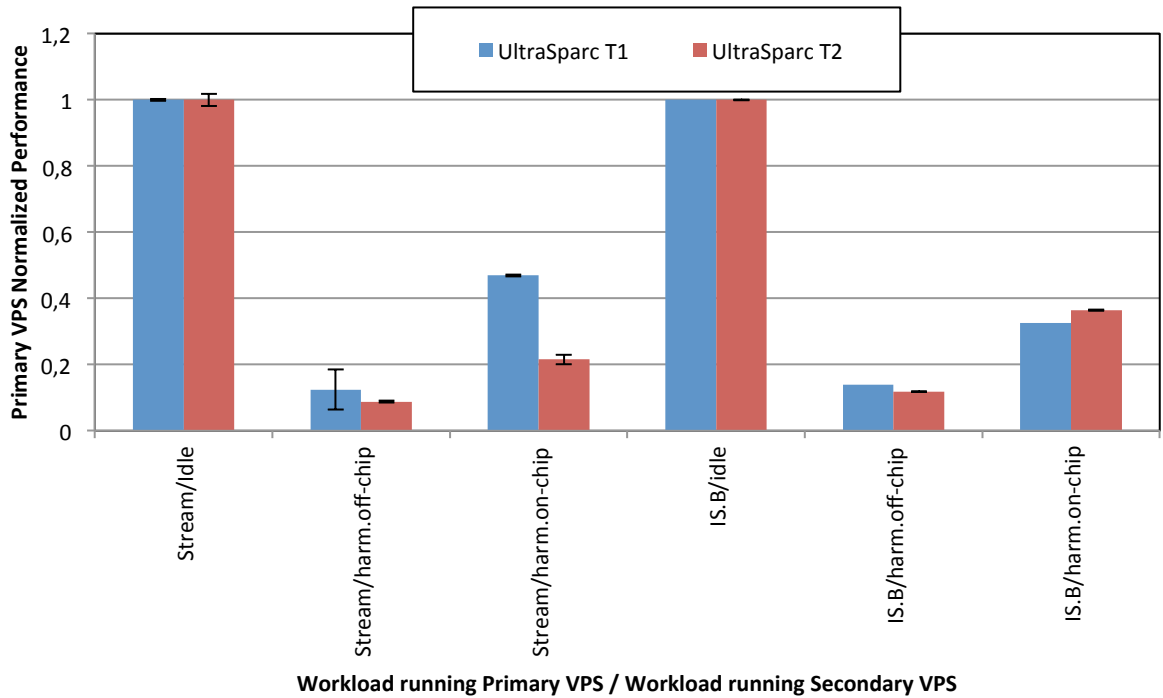

***Fig. 2.*** *Performance interference induced by secondary VPS running different harmful application variations on Sun Fire T1000 and Sun Enterprise T5120*

To determine the actual off-chip interference of the two kernels we measured the performance effects of both configurations and the bandwidth available to the primary VPS running STREAM [12] on it. STREAM is a synthetic benchmark which measures the memory performance of four long vector operations. The results, obtained with a confidence interval of 95%, are shown in **Fig. 2**. As can be observed, *harm.off-chip* consumes up to 88% of the off-chip bandwidth in UltraSparc T1 and 91% in UltraSparc T2. In contrast, *harm.on-chip* reduces the off-chip bandwidth available to the primary VPS by 55% in UltraSparc T1 and 77% in UltraSparc T2. It is unavoidable that *harm.on-chip* has to generate L2 misses to evict the primary VPS blocks; therefore it consumes some off-chip bandwidth. The UltraSparc T2 system is more sensitive to the harmful application because the misbehaving application is able to use a higher share of the off-chip bandwidth. Therefore, there is less L2 contention because of the higher number of L2 banks and this favors more off-chip access for the secondary VPS in this system.

To observe the effects of each synthetic workload influence on primary VPS performance we chose an easy to understand multithread application running on it. We selected the IS benchmark of the NAS Parallel Benchmarks [7] version OpenMP 3.2. The application applies the bucket sort algorithm over a given number of 32-bit integers using 10 buckets. The application is composed of ten iterations where, for the most part, each thread is passing sequentially through arrays. Therefore, depending on the problem size, the application will be sensitive to off-chip bandwidth or on-chip resources.

The configuration for the off-chip limit will be $2^{25}$ integers with 32-bit integers (denoted by IS.B). Although IS is very simple, it is easy to understand and additionally overcomes the single FPU limitation of UltraSparc T1. As illustrated in **Fig. 2**, similarly to the *Stream* scenario *harm*.o*ff-chip* is the most damaging micro-benchmark in both systems.

To sum up, it is clear that both micro-kernels have a significant impact on this system running a paper-and-pencil application such as IS in the primary VPS. Off-chip bandwidth emerges as the most critical resource in UltraSparc T1/T2 processors.

### 3.2 Intel Xeon X5472 Harpertown

In this architecture we will use three different configurations for the primary and secondary VPSs. In the first one, the primary and secondary VPSs share a die, both being restricted to using only one core each (light and medium grey in Fig. 1b). In the second configuration, each VPS has a full die to its disposal, both running in the same chip. Looking at Fig. 1b, the primary zone will run in the light and medium grey cores and the secondary in the dark ones. In the third configuration, designated as worst-case scenario, each VPS has 2 cores but interleaved in the chip. Note that this situation is easily solvable by simply making an adequate VPS core allocation. As a word of warning, by default, the Linux kernel used in the system interleaves consecutive core ids.

In order to keep the results clear, for each VPS combination we will only show the synthetic workload that introduces the greatest performance degradation. In the first VPS combination, the offending workload is configured with one thread per available physical core in the VPS[3] (one thread in the first configuration, two threads in the others), a private per-thread array of 64 MBytes and a stride of 64 bytes. The other combinations use a stride of 256KBytes. As the L2 has 4096 sets, this stride makes references to the same set in L2 and, thus it will mainly degrade off-chip bandwidth. **Fig. 3** shows the performance of IS.C (sorting $2^{27}$ 32-bit integers) observed for each combination of primary and secondary VPS. Similarly to the case of UltraSparc processors, interference at on-chip level is less critical than off-chip bandwidth.

The parameters chosen for the harmful workload have a significant impact on any combination which generates interference in off-chip memory controller access. Surprisingly, not interleaving the distribution of VPS does not mitigate the problem. In this computer, the off-chip memory subsystem design makes the Memory Controller Hub (MCH) a critical component, easily flooded by the offending VPS. Note that having more cores in the offending VPS just makes things worse.

Chip-interleaved is the worst-case combination, because the secondary VPS not only consumes off-chip bandwidth but also pollutes the on-chip shared L2 in both cores assigned to the primary VPS.

As expected, when the primary and secondary VPSs are sharing the L2 cache, the performance impact of using a stride of 64 bytes was the most harmful. Under these circumstances the micro-kernel pollutes the cache, increasing the miss ratio of the primary VPS application.

---

[3] The processor has no SMT support

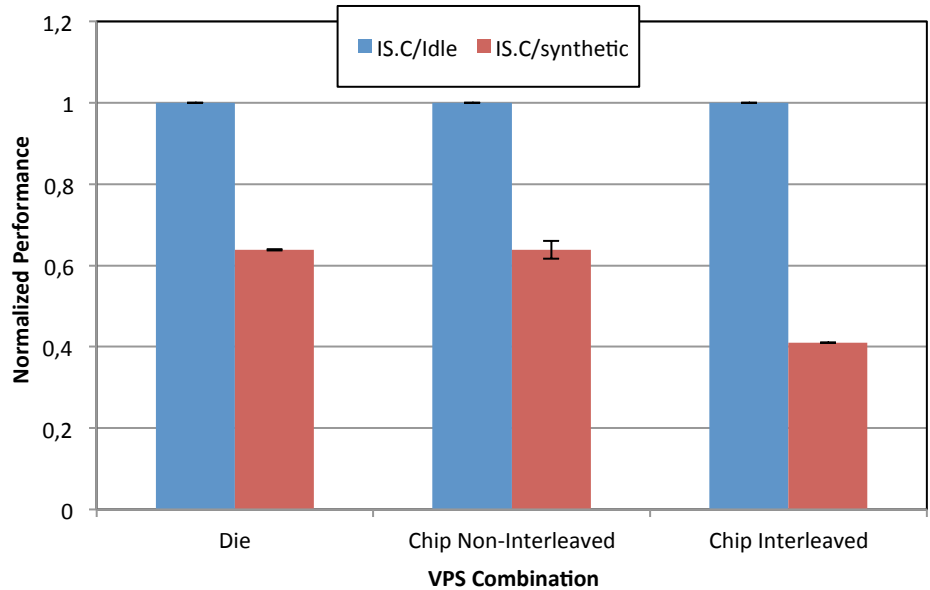

*Fig. 3.* HP Proliant DL160 IS.C performance degradation running worst-case synthetic workload

In general, the performance effects are less appreciable in this system than in the UltraSparcs. The higher capability of the processor and memory hierarchy limit the adverse effects of the synthetic workload. Given the complexity and the scarcity of the available information, it is difficult to present and validate another hypothesis. In any case, the effects are clear, achieving only 40% of the original performance in the worst-case.

## 4  Performance Isolation in Consolidated Servers

### 4.1  Sun Fire T1000 and Sun Enterprise T5120

To determine the impact of a realistic workload we run SpecWeb2005 on the primary VPS, with the synthetic workloads proposed in Section 3.1 running on the secondary VPS. Now, we will analyze how a consolidated server reacts when the offending application is virtualized in the same CMP. It might appear to be seen as an extreme situation, but it is possible using the currently available isolation techniques.

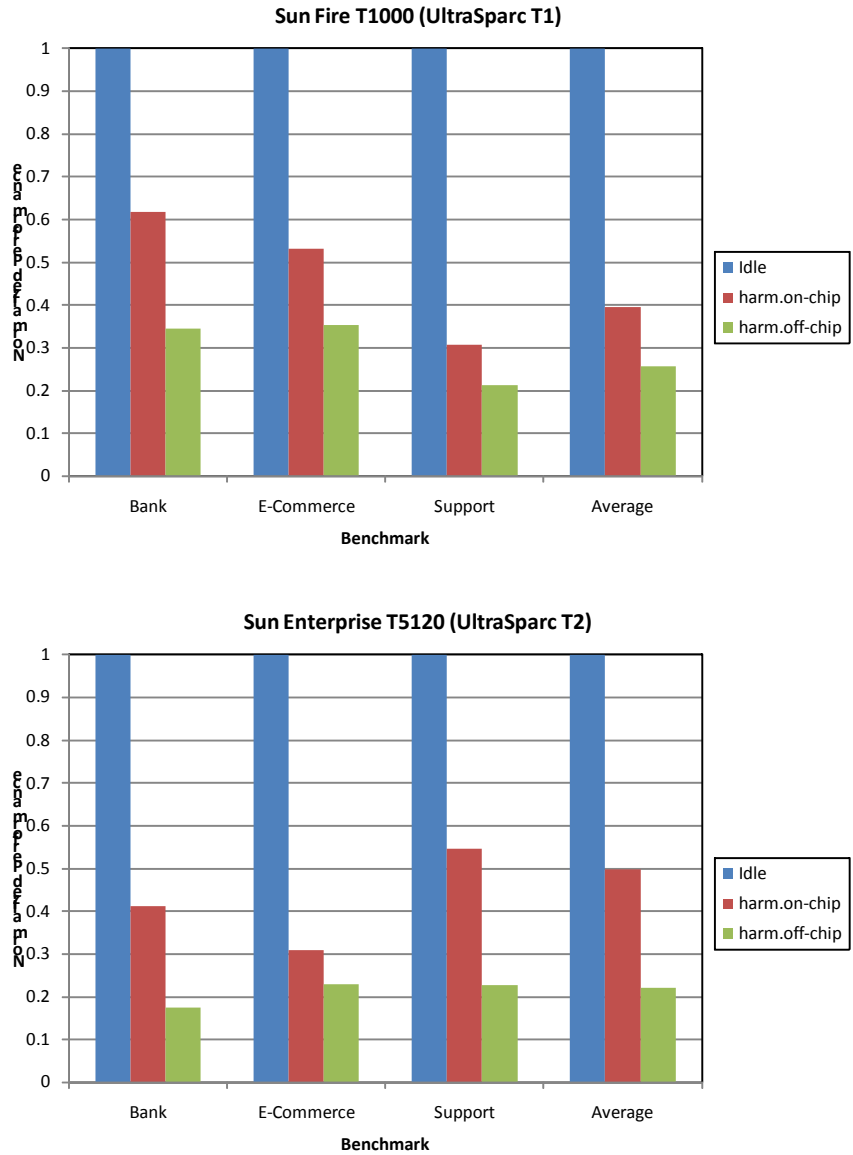

*Fig. 4.* *Performance of SPECWeb2005 running in primary VPS. It is normalized with an idle secondary VPS. (above) Sun Fire T1000 - UltraSparc T1, (below) Sun Enterprise T5120 - UltraSparc T2.*

The performance results for both systems are shown in **Fig. 4**. In UltraSparc T1 when the other VPS is executing the synthetic workloads, the performance drops significantly. In the worst case, *Support* can handle almost 5 times fewer clients when the secondary VPS is executing *harm.off-chip*. Most of the requests that fail to complete in TIME_TOLERABLE (5 seconds in this benchmark) are requests for the *product*[4] page. This request is the one that needs to execute most instructions[5]. The performance reduction of *E-commerce* and *Bank*, although significantly affected, is less. Probably, the requirement for data encryption implies a higher locality for the executed code, so it has a smaller impact on L2 and off-chip bandwidth.

The bank workload is usually limited by the number of simultaneous sessions a server can handle when other applications are running in the other VPS. When running alone or with the same benchmark in the other Solaris zone, most of the failed connections happen in the *login* page. If a client successfully logs in, it will complete most of the bank operations in time. The difference is that, when the other VPS is idle, the server can handle 2.83X more users than when the other VPS is running the harmful micro-kernel. Something similar happens with *E-commerce*. Note that the number of cores available to the primary VPS and the secondary VPS is fixed, regardless of whether the harmful micro-benchmark is running or not.

In the case of UltraSparc T2, on average, the performance impact is slightly higher. This is coherent with the behavior observed in **Fig.2**. For the system evaluated, it seems that the off-chip access regulation is even more necessary than in UltraSparc T1. In the *Bank* benchmark when the secondary VPS is idle, it is capable of serving 5.7X more clients than when it is running the off-chip bandwidth consuming benchmark, which constitutes a dangerous performance loss.

As we can see in **Fig. 4**, *harm.off-chip* is worse than *harm.on-chip*. In the UltraSparc T1 we analyzed a snapshot of the execution using *Cputrack* from the Solaris Performance Tools, to collect L2 data from the processor performance counters in UltraSparc T1. As can be seen in **Fig. 5** the application by itself has a high L2 miss per thousand instructions (MPKI). Most of those misses correspond to instructions, indicating the large code footprint for the benchmark. The misses added by *harm.on-chip* workload are clearly visible but they do not affect performance to the same extent. In fact, the performance degradation follows the opposite trend. The observed behavior reinforces the idea that off-chip bandwidth is the most valuable asset in the CMP.

---

[4] It is a request type of this particular benchmark, such as *download, file,*

[5] Note that this suggests that the benchmark is CPU-bounded. Had it been network-bounded or disk-bounded, the failed requests would have been at *download* or *file*, whose responses are much larger.

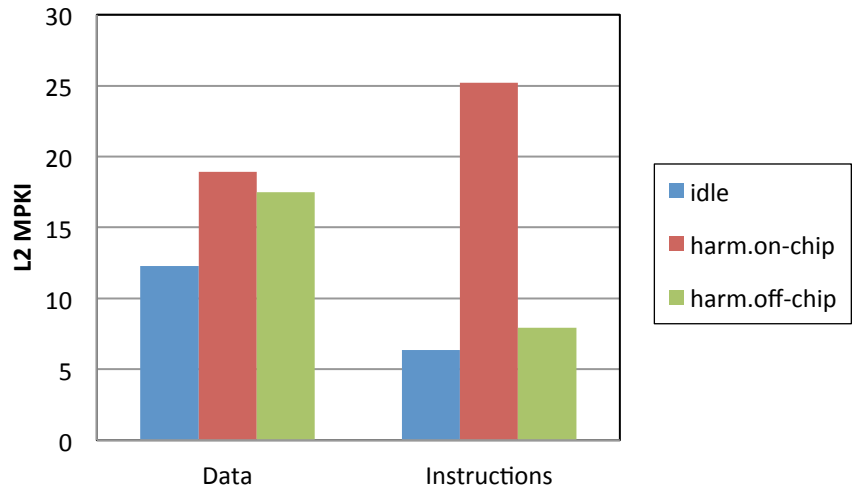

***Fig. 5.*** *Average, L2 Misses per thousand Instructions of SPECweb2005 Bank when running alone or with the harm.on-chip and harm.off-chip synthetic workloads.*

## 4.2 HP Proliant DL160 G5

With this hardware platform, we run the SPECweb2005 workload against the worst-case synthetic workloads analyzed in Section 3.2. We analyzed the same 3 VPS configurations:

- Two VPSs sharing the same die (Fig.1b light and medium grey).
- The two VPSs sharing the chip, each running in a different die. In Fig. 1b, the primary VPS runs in the light and medium grey cores and the secondary runs in the dark grey ones.
- The two VPSs in the same chip, running in consecutive cores. In practice, this means that each VPS uses one core of each die in chip 0.

**Fig. 6** shows the performance degradation of the benchmark running the synthetic workload in the other VPSs. The *support* benchmark consists of non-encrypted traffic, so analyzing this benchmark we can account for the performance loss caused by running the JSP scripts. Moreover, 7% of the responses of this benchmark are static, so a minimal amount of CPU is required to solve them (TCP overhead, tomcat parsing the request). This is the reason why *support* is the benchmark that is most resilient to the harmful workload.

The *bank* benchmark is completely composed of SSL traffic, while *e-commerce* has a mix of encrypted and plain text requests. So, in these benchmarks, the CPU-intensive part comprises the execution of the JSP as well as the encryption of the traffic. As **Fig. 6** illustrates, L2 cache locality is critical for these SSL benchmarks. The "same die" and "same chip" interleaved configurations, which share the L2 with the other VPSs, suffer significant degradation in these benchmarks. This is especially true for *bank*, which, as stated before, uses more encrypted traffic than *e-commerce*. The same chip configuration, which depends mainly on off-chip bandwidth, also sees some performance drop in these benchmarks.

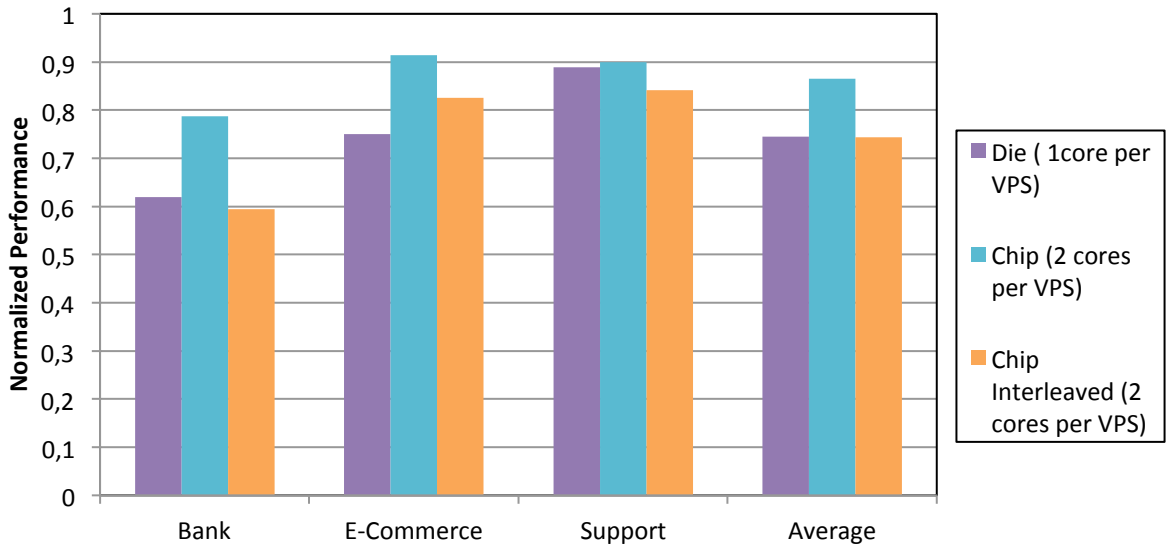

***Fig. 6.*** *Performance of SPECWeb2005 running in the primary VPS while the synthetic benchmark runs in the secondary VPS. The performance is normalized with the performance of SPECWeb2005 when the secondary VPS is idle.*

In contrast to what happens in the UltraSparc processors, in this architecture the effect of the harmful application running in secondary VPS is less relevant. On average there is a performance loss of only 12% when running the VPS in separate dies. We can appreciate that the offending VPS can clog primary VPS off-chip access. Nevertheless, when the VPSs are sharing L2 cache the effects are greater with up to a 40% performance loss in some phases of the benchmark. Probably, the significantly larger Xeon on-chip caches make the application less sensitive to off-chip bandwidth availability. In any case, the architecture is very complex, very different and, moreover, the information about the Xeon and the chipset used is too scarce to extract any conclusions about the differences in performance between them.

## 5    Related work

The notion of Quality of Service in CMPs (CQoS) was originally introduced by Iyer in [5]. This work advocates providing some sort of hardware mechanism to limit the performance unpredictability induced by CMP characteristics. Iyer et al extended it in [6], where the authors use simulations to evaluate different QoS policies to control resources or performance. However, their work does not analyze the interferences that lead to the performance degradation. They focus on the amount of resources devoted to each application. Surprisingly, and contrary to what we have shown in this paper, they pay more attention to the cache space than to memory bandwidth. In [19], the authors investigate low-overhead mechanisms for fine-grain monitoring of the use of shared cache resources proposing the CacheScouts monitoring architecture. In [17] Padala et al. describe a framework to control coarse-grain resources (CPU and disk) and present their results with real hardware. However, as they are measuring CPU utilization as reported from the Xen Manager, their approach would not be able to spot the off-chip bandwidth bottleneck and act upon it.

In [15] Nesbit et al introduce the notion of Virtual Private Machine framework. This abstraction allows the software to adjust the hardware policies to its requirements. The proposals suggest a mechanism to keep shared resources under control throughout the memory hierarchy of CMP systems. In particular [13] proposes a solution to regulate off-chip bandwidth through fair memory scheduling. In [14] a method to fairly partition the on-chip bandwidth utilization and last level cache was introduced.

Thread interference in caches has been widely studied in the literature. Kim et al [9] proposed fair partitioning of the cache in order to prevent cache trashing while Hsu et al [4] considered different performance targets for cache partitioning. Software solutions have also been tried. In [11] Kannan et al aim to improve the performance of an individual application at the cost of the potential detriment of others with guidance from the operating environment. We are not aware of any real-world evaluation of QoS in CMPs.

Marty and Hill [10] proposed virtual hierarchies to manage Virtual Machines (VM) in multi-core tiled architectures. Instead of having a big shared L2 cache as current processors have, they advocate private per-core L2s. The paper shows the cache coherence problems of having so many cores and proposes virtual hierarchies to reduce unnecessary coherent traffic when the hardware knows it is running many VMs.

From our perspective, most of the related work is composed of studies that analyze or suggest solutions from a purely architectural point of view using simulation tools. Those tools make it essential to cut down the workloads and simplify the setup. In contrast, in this work we try to use a completely realistic setup for the hardware and software environment to assert that Chip QoS is a top priority issue.

It should be noted that in other works, such as [2], the problem is faced from the operating system side. In any case, using the hardware available in current systems [18], the benefit is limited.

## 6     Conclusions and Future Work

In this paper we have evaluated a real world scenario in which Virtual Private Servers running on state-of-the-art CMP systems cannot guarantee achievable performance. Quality-of-Service guarantees are important for the success of commercial cloud computing, but with current technology, a service provider cannot realistically sign a Service Level Agreement with its customers. Under those circumstances the performance of a given VPS depends on what is running in other VPSs, thus failing to provide performance isolation and guarantee Quality of Service. Not surprisingly, even with the reduced number of cores per chip, the most valuable resource in the systems under study is the off-chip bandwidth. Increasing the number of cores per chip will only exacerbate this problem.

This lack of Quality of Service provision could be a serious issue for the ubiquity of next generation multi-core CMPs. With multi-core CMPs, in order to maintain cost and power efficiency of computing infrastructures, it will be necessary to consolidate different services in a single chip or system.

From our perspective, it is essential to provide some mechanisms to enable the software layer to apply the necessary policies to guarantee the desired level of performance. Given the fine-grained interference at intra-chip level, we believe that software-only approaches will not be enough to provide a good level of QoS. Only with the proper hardware facilities and their corresponding software interface, will we be able to solve the problem.